\documentclass[USletter,10pt]{article}
\pdfoutput=1
\usepackage{jheppub}

\usepackage{alltt}
\usepackage{amsfonts}
\usepackage{amsmath}
\usepackage{amssymb}
\usepackage{amsthm}
\usepackage{array}
\usepackage[USenglish]{babel}
\usepackage{breakcites}
\usepackage[font=footnotesize,labelfont=bf]{caption}
\usepackage{comment}
\usepackage{csquotes}
\usepackage{dsfont}
\usepackage{enumerate}
\usepackage{enumitem}
\usepackage[bottom]{footmisc}	%Bottom figure, but above footnotes
\usepackage{braket}
\usepackage[T1]{fontenc}
\usepackage{graphicx}
\usepackage{hyperref}
\usepackage{imakeidx}
\usepackage[utf8]{inputenc}
\usepackage[USenglish]{isodate}
\usepackage{lmodern}
\usepackage{mathdots}
\usepackage{mathrsfs}
\usepackage{mathtools}
\usepackage{mdframed}
\usepackage{microtype} %Breaks lines in references, do not work for ISBN
\usepackage{multicol}
\usepackage{multirow}
\usepackage{nameref}
\usepackage{ragged2e}
\usepackage{physics}
\usepackage{setspace}
\usepackage{siunitx}
\usepackage{slashed}
\usepackage{stackengine}
\usepackage{subcaption}
\usepackage{tasks}
\usepackage{tensor}
\usepackage[titles]{tocloft}
\usepackage{upgreek}
\usepackage{xfrac}
\usepackage{verbatim}
\usepackage{wrapfig}

\DeclareMathAlphabet\EuScript{U}{eus}{m}{n}
\SetMathAlphabet\EuScript{bold}{U}{eus}{b}{n}

\setenumerate{font=\itshape}

\renewcommand{\mkbegdispquote}[2]{\itshape}

\makeatletter
\def\@fpheader{~}
\makeatother

\makeatletter
\newcommand*{\currentname}{\@currentlabelname}
\makeatother

\makeatletter
\@addtoreset{equation}{section}
\makeatother
\numberwithin{equation}{section}

\makeindex[intoc, title=Index]
\newcommand {\pd}[2][ ]{
  \ifx #1 { }
    \frac{\partial}{\partial #2}
  \else
    \frac{\partial^{#1}}{\partial #2^{#1}}
  \fi
}
\newcommand {\pdo}[3][ ]{
  \ifx #1 { }
	\frac{\partial #2}{\partial #3}
  \else
	\frac{\partial^{#1} #2}{\partial #3^{#1}}
  \fi
}

\newenvironment{aside}
	{
	\begin{mdframed}[style=0,%
		leftline=false,rightline=false,leftmargin=2em,rightmargin=2em,%
		innerleftmargin=0pt,innerrightmargin=0pt,linewidth=0.75pt,%
		skipabove=12pt,skipbelow=20pt]\small}
	{\end{mdframed}
	}

\newtheoremstyle{mytheoremstyle} % name
{.4\baselineskip\@plus.2\baselineskip\@minus.2\baselineskip}%{\topsep}                    % Space above
{.4\baselineskip\@plus.2\baselineskip\@minus.2\baselineskip}%{\topsep}                    % Space below
{\itshape}                   % Body font
{}                           % Indent amount
{\bfseries}                   % Theorem head font \scshape
{.}                          % Punctuation after theorem head
{.5em}                       % Space after theorem head
{}  % Theorem head spec (can be left empty, meaning ‘normal’)

\theoremstyle{mytheoremstyle}



\newcommand{\IGNORE}[1]{}

\renewcommand{\vec}[1]{\boldsymbol{\mathbf{#1}}}

\DeclareFontFamily{U}{FdSymbolA}{}
\DeclareFontShape{U}{FdSymbolA}{m}{n}{
	<-> s*[.4] FdSymbolA-Regular
}{}
\DeclareSymbolFont{fdsymbol}{U}{FdSymbolA}{m}{n}
\DeclareMathSymbol{\smallcirc}{\mathord}{fdsymbol}{"60}

\makeatletter
\newcommand{\hollowcolon}{\mathpalette\hollow@colon\relax}
\newcommand{\hollow@colon}[2]{%
	\mspace{1mu}%
	\vbox{%
		\hbox{$\m@th#1\smallcirc$}
		\nointerlineskip
		\kern.45ex
		\hbox{$\m@th#1\smallcirc$}
		\kern-.06ex
	}%
	\mspace{1mu}%
}
\makeatother

\makeatletter
\DeclareFontFamily{U}{tipa}{}
\DeclareFontShape{U}{tipa}{m}{n}{<->tipa10}{}
\newcommand{\arc@char}{{\usefont{U}{tipa}{m}{n}\symbol{62}}}%

\newcommand{\arc}[1]{\mathpalette\arc@arc{#1}}

\newcommand{\arc@arc}[2]{%
	\sbox0{$\m@th#1#2$}%
	\vbox{
		\hbox{\resizebox{\wd0}{\height}{\arc@char}}
		\nointerlineskip
		\box0
	}%
}
\makeatother

\newcounter{innerlist} 
\renewcommand{\theinnerlist}{\itshape{\arabic{innerlist}}} 
\newcommand{\inneritem}[1]{\refstepcounter{innerlist}\theinnerlist.\ #1}
\newenvironment{rowenumerate}[1]{\setcounter{innerlist}{0} \begin{tabular}{#1}}{\end{tabular}}

\newcolumntype{L}[1]{>{\raggedright\let\newline\\\arraybackslash\hspace{0pt}}m{#1}}
\newcolumntype{C}[1]{>{\centering\let\newline\\\arraybackslash\hspace{0pt}}m{#1}}
\newcolumntype{R}[1]{>{\raggedleft\let\newline\\\arraybackslash\hspace{0pt}}m{#1}}
\newcolumntype{J}[1]{>{\justify\let\newline\\\arraybackslash\hspace{0pt}}m{#1}}

\newcommand{\x}{\hspace{.8pt}}

\title{\boldmath The Holographic Entropy Cone for Five Regions}

\author{Sergio Hern\'{a}ndez Cuenca}

\affiliation{Department of Physics, University of California Santa Barbara, California 93106, USA}

\emailAdd{sergiohc@physics.ucsb.edu}

\abstract{
Even though little is known about the quantum entropy cone for $N\geq4$ subsystems, holographic techniques allow one to get a handle on the subspace of entropy vectors corresponding to states with gravity duals. For static spacetimes and $N$ boundary subsystems, this space is a convex polyhedral cone known as the holographic entropy cone $\mathcal{C}_N$ for $N$ regions. While an explicit description of $\mathcal{C}_N$ was accomplished for all $N\leq4$ in the initial study, the information given about larger $N$ was only partial already for $\mathcal{C}_5$. This paper provides a complete construction of $\mathcal{C}_5$ by exhibiting graph models for every extreme ray orbit generating the cone defined by all proven holographic entropy inequalities for $N=5$. The question of whether there exist additional inequalities for $5$ parties is thus settled with a negative answer. The conjecture that $\mathcal{C}_5$ coincides with the analogous cone for dynamical spacetimes is supported by demonstrating that the information quantities defining its facets are primitive.
}

\makeindex

\begin{document}
	
\maketitle
\flushbottom

\pagenumbering{roman}

%Addapting to page number
%\setstretch{1.14} %Inter-line spacing
%\setlength{\abovedisplayskip}{5pt} %Equation display spacing above
%\setlength{\belowdisplayskip}{5pt} %Equation display spacing above

%\newpage
\pagenumbering{arabic}

\section{Introduction}

The most broadly studied limit of the AdS/CFT correspondence conjectures a holographic duality between certain strongly coupled gauge theories and classical gravity~\cite{Maldacena1999,Aharony1999a}. More explicitly, for a pair of gauge-gravity dual theories, the AdS/CFT dictionary poses a specific spacetime geometry as the gravitational counterpart of a given state in the Hilbert space of the quantum theory. On grounds of such a duality, it is of interest to determine which quantum states are dual to classical bulk geometries. A remarkable finding from the study of holographic entanglement is that, regardless of the theory, quantum states with particular patterns of correlations do not admit smooth geometric duals~\cite{Hayden2013,Balasubramanian2014}.

At the heart of this result lies the Ryu-Takayanagi (RT) proposal, which states that for static bulk geometries the entanglement entropy $S_{\textrm{A}}$ of a spatial region $A$ of the boundary conformal field theory is given by~\cite{Ryu2006,Nishioka2009,Rangamani2017}
\begin{equation}\label{RTformula}
	S_A = \min_\mathcal{A} \; \frac{\textrm{area } \mathcal{A}}{4 G_\mathrm{N}},
\end{equation}
where the minimization is performed over all bulk codimension-$2$ surfaces homologous to $A$ and such that $\partial A = \partial \mathcal{A}$. The Hubeny-Rangamani-Takayanagi (HRT) prescription gives the covariant generalization of RT that applies to arbitrary dynamical spacetimes~\cite{Hubeny2007,Rangamani2017}. That the RT formula should reproduce the results of the von Neumann entropy for arbitrary partitions of a quantum state establishes a necessary condition for the existence of a smooth bulk dual. The discovery that there exist valid holographic entropy inequalities which are not true in quantum theory means that this necessary condition is not met by arbitrary quantum states. In particular, this is the case for the inequality known as monogamy of mutual information (MMI)~\cite{Hayden2013},
\begin{equation}\label{key}
	I_2(\textrm{A:BC}) \geq I_2(\textrm{A:B}) + I_2(\textrm{A:C}),
\end{equation}
defined here in terms of the mutual information $I_2(\textrm{A:B})=S_{\textrm{A}}+S_{\textrm{B}}-S_{\textrm{AB}}$, and where $\textrm{A}$, $\textrm{B}$ and $\textrm{C}$ stand for three disjoint regions. This inequality has been proven true holographically for arbitrary dynamical spacetimes~\cite{Hayden2013,Wall2014}, yet is easily violated quantum mechanically (e.g. by the GHZ state).

It follows that characterizing entanglement properties via entropy inequalities provides a powerful criterion to determine whether a quantum state can possibly be geometric (i.e. whether it can be holographically dual to a classical geometry). The formalization of this idea was carried out in~\cite{Bao2015}, which introduced what is known as the holographic entropy cone to parametrize the space of allowed entropies for geometric states. The purpose of this work is to continue the systematic study and enumeration of holographic entropy inequalities. The picture for $4$ and fewer regions was completed in~\cite{Bao2015}, where partial results were also found for $5$ regions. Building on this previous work, the complete construction of the holographic entropy cone for $5$ regions is produced here. A direct corollary is that no additional entropy inequalities are needed for the completion of the $5$-party cone.

\section{Framework and Approach}

Let $\Sigma$ be a spacelike slice of the spacetime manifold of a quantum field theory in a state which admits a holographic description in terms of a smooth bulk geometry. Consider $N\in\mathbb{Z}^+$ arbitrary nonempty codimension-$1$ disjoint subsets $X_i \subset \Sigma$, where $i\in [N] \equiv \{ 1,\dots,N\}$. Any such $X_i$ will be referred to as a monochromatic region of colour $i$. One also defines the set of polychromatic indices $\wp_N$ as the power set of $[N]$ with the empty set removed. The latter has cardinality $D=2^N-1$ and its elements $I\in\wp_N$ are used to label polychromatic regions $X_I \equiv \bigcup_{i\in I} X_i$. Denoting the entanglement entropy of each region $X_I$ by $S_I$, one may construct a $D$-tuple $ \vec{S} \equiv \left\{ S_I \, \mid \, I \in \wp_N \right\}$. Canonically ordering its entries by increasing cardinality of $I$ and then lexicographically, $\vec{S} \in \mathbb{R}^D$ defines an entropy vector.

Every entropy $S_I$ of a collection of regions can be computed holographically to leading order in the central charge of the boundary theory using the HRT prescription~\cite{Hubeny2007}. For static bulk geometries for which this construct reduces to the RT formula \eqref{RTformula}~\cite{Ryu2006}, the space $\mathcal{C}_N \subset \mathbb{R}^D$ of all physically realizable holographic entropy vectors $\vec{S} \in \mathbb{R}^D$ is known as the holographic entropy cone $\mathcal{C}_N$ for $N$ regions. It was shown in \cite{Bao2015} that this space is indeed a convex cone which is closed, rational and polyhedral. The Farkas-Minkowski-Weyl theorem recasts polyhedrality into the existence of two dual representations of such convex cones~\cite{Grunbaum2003a}:
\begin{itemize}%[itemsep=0pt,leftmargin=14pt]
%	\vspace{-8pt}
	\item Facet representation: $\mathcal{C}_N$ can be constructed as the intersection of a finite number of half-spaces specified by entropy inequalities of the form $\vec{S} \cdot \vec{Q}_j \geq0$, where $\vec{Q}_j\in\mathbb{R}^D$. The minimal collection $\{ \vec{Q}_j \in \mathbb{R}^D \}$ of such vectors is unique and geometrically defines the support hyperplanes or facets of the cone.
	\item Extreme ray representation: $\mathcal{C}_N$ can be finitely generated as the conical hull of a set of vectors. The minimal collection of such vectors is unique and consists of the extreme rays $\{ \vec{e}_k \in \mathcal{C}_N \}$ of the cone, i.e. the vectors in $\mathcal{C}_N$ which cannot be conically spanned by other vectors in $\mathcal{C}_N$.
%	\vspace{-6pt}
\end{itemize}
Importantly, since $\mathcal{C}_N$ is closed, for every extreme ray $\vec{e}_k\in\mathcal{C}_N$ there exists a bulk geometry and a choice of boundary regions such that their corresponding entropy vector $ \vec{S} \propto \vec{e}_k $~\cite{Bao2015}. Also, by virtue of being rational, the facet vectors and extreme rays of $\mathcal{C}_N$ can be written with integer coordinates. In particular, by the non-negativity of entanglement entropy, every $\vec{e}_k$ has non-negative integer entries, as will be seen.

Constructing the holographic entropy cone $\mathcal{C}_N$ for $N$ regions amounts to finding a representation of it. A collection of proven entropy inequalities for $N$ parties does not necessarily provide a complete representation of $\mathcal{C}_N$. More specifically, supposing that such collection of inequalities represents a cone $\tilde{\mathcal{C}}_N$, that they are true entropy inequalities only guarantees that $\mathcal{C}_N \subseteq \tilde{\mathcal{C}}_N$. Proving that the facets of the two cones in fact coincide is better done in the dual description in terms of extreme rays. In particular, if for every extreme ray of $\tilde{\mathcal{C}}_N$ one is able to find a geometry whose entropy vector lies on it, then convexity immediately implies that $\mathcal{C}_N = \tilde{\mathcal{C}}_N$.

This strategy was implemented in \cite{Bao2015} to construct the holographic entropy cones for $N\leq4$. For $N=5$, the authors successfully found and proved by contraction five new entropy inequalities, but left as an open question whether this set was complete. A thorough understanding of $\mathcal{C}_5$ has thus been lacking. In this work, the complete representation of the holographic entropy cone $\mathcal{C}_5$ for $5$ regions is provided by explicit construction of its extreme rays. One of the outcomes is that there are no new holographic inequalities for $5$ parties, so that the facets of $\mathcal{C}_5$ are precisely certain upliftings of known inequalities for $N\leq3$ and the five new ones proven in \cite{Bao2015}.

The construction of the extreme rays of $\mathcal{C}_5$ is given here in terms of graph models as introduced in \cite{Bao2015}. The key theorem behind this combinatorial approach is that $\vec{S} \in \mathcal{C}_N$ if and only if there exists a graph model that realizes $\vec{S}$. In other words, the holographic entropy cone and the analogously defined graph-model entropy cone are identical. A graph model for $N$ parties is an undirected graph $(V,E)$ with $V$ vertices and $E$ edges, where a subset $\partial V \subseteq V$ is coloured by a map $c : \partial V \to [N]$. As the nomenclature suggests, a vertex coloured by $i$ stands as the graph representative of the monochromatic region $X_i$ in the boundary theory. The elements of $\partial V$ are thus called boundary vertices, while those in the complement $V\smallsetminus\partial V$ are called bulk vertices. Edges are assigned non-negative edge capacities by a weight map $E \to \mathbb{R}_{\geq0}$. Then, the entropy $S_I$ of a polychromatic subset of boundary vertices $ \partial V_I \equiv c^{-1}[I] \subset \partial V$ is given by the maximum flow between multisources $V_I$ and multitargets $ \partial V \smallsetminus \partial V_I $ which respects the edge capacities. By the max-flow min-cut theorem, this is equivalent to the prescription that defines $S_I$ as the total weight in the minimum cut which disconnects source from sink. Physically, the latter is equivalent to the RT prescription, while the former corresponds to the bit-thread formulation of entanglement~\cite{Freedman2017}.

\section{The Holographic Entropy Cone for Five Regions}

The action of the symmetric group $S_N$ which relabels the regions $X_i$ clearly leaves $\mathcal{C}_N$ invariant. This symmetry extends to an $S_{N+1}$ symmetry which implements the exchange of any $X_i$ with the purifier $O \equiv \Sigma\smallsetminus \bigcup_{i\in [N]} X_i$. Henceforth, statements about symmetries refer to the extended symmetry group $S_{N+1}$. The following subsections detail the description of $\mathcal{C}_5$ in its two representations.

\subsection{Facets}

The starting point of the strategy described above is a set of true inequalities for $N=5$ which is to be proven complete. This set consists of $372$ inequalities, which reduce to just $8$ when quotiented by symmetry. Table \ref{inequalities} shows a representative inequality for each symmetry orbit\footnote{For the explicit construction of $N=5$ objects, monochromatic indices are assigned alphabetic values rather than positive integers, and polychromatic indices are written by juxtaposition of letters rather than tuples.}. The first three are upliftings of well-known inequalities for $N\leq3$, whereas the last five are new to $N=5$. Inequality \ref{SA} is the trivial uplifting of subadditivity, whose orbit includes instances of the Araki-Lieb inequality too. Inequalities \ref{MMI1} and \ref{MMI2} are two different upliftings of MMI, which can be more compactly written in terms of the tripartite information as $I_3(\textrm{A:B:C})\leq0$ and $I_3(\textrm{A:BC:DE})\leq0$, respectively. Inequality \ref{QCyclic} is the $5$-party instance of an infinite family of cyclic entropy inequalities\footnote{In the literature, this inequality is often quoted with manifest cyclicity of the subsystem labels. For consistency with the notation used for the other inequalities, symbols and their labels are ordered lexicographically here.}~\cite{Bao2015}. Like inequality \ref{QCyclic}, the remaining four were proven by contraction for the RT case in \cite{Bao2015}. This set of inequalities defines a cone in entropy space which will be shown to be precisely the holographic entropy cone $\mathcal{C}_5$ for $5$ regions in the next section. A natural question, however, is how to arrive at these inequalities, in particular the last five, in the first place. As for now, only \cite{Hubeny2018b} succeeded in algebraically deriving these as candidate inequalities using the formalism of the holographic entropy arrangement~\cite{Hubeny2018a,Hubeny2018b}.

\begin{table}
\caption{Representatives for each of the $8$ inequality orbits of the holographic entropy cone $\mathcal{C}_5$ for $5$ regions. Respectively, their orbit lengths are $15$, $20$, $45$, $72$, $10$, $60$, $60$ and $90$, thus defining $372$ facets for $\mathcal{C}_5$ in a $31$-dimensional entropy space.}
\label{inequalities}
\setlength{\tabcolsep}{6pt}
\renewcommand{\arraystretch}{.6}
\noindent
\fontsize{10}{11} \selectfont
\hrule
\begin{rowenumerate}{J{0.47\textwidth} J{0.47\textwidth} }
	\inneritem $ \label{SA}	S_{\textrm{A}} + S_{\textrm{B}} \geq S_{\textrm{AB}} $ &
	\inneritem $\label{MMI1}	S_{\textrm{AB}} + S_{\textrm{AC}} + S_{\textrm{BC}} \geq S_{\textrm{A}} + S_{\textrm{B}} + S_{\textrm{C}} + S_{\textrm{ABC}}$ \\
	\inneritem $\label{MMI2}	S_{\textrm{ABC}} + S_{\textrm{ADE}} + S_{\textrm{BCDE}} \geq S_{\textrm{A}} + S_{\textrm{BC}} + S_{\textrm{DE}} + S_{\textrm{ABCDE}}$ &
	\inneritem $\label{QCyclic}		S_{\textrm{ABC}} + S_{\textrm{ABD}} + S_{\textrm{ACE}} + S_{\textrm{BDE}} + S_{\textrm{CDE}} \geq S_{\textrm{AB}} + S_{\textrm{AC}} + S_{\textrm{BD}} + S_{\textrm{CE}} + S_{\textrm{DE}} + S_{\textrm{ABCDE}}$ \\
	\inneritem $\label{Q2}	S_{\textrm{ABC}}+S_{\textrm{ABD}}+S_{\textrm{ABE}}+S_{\textrm{ACD}}+S_{\textrm{ACE}}+S_{\textrm{ADE}}+S_{\textrm{BCE}}+S_{\textrm{BDE}}+S_{\textrm{CDE}} \geq S_{\textrm{AB}}+S_{\textrm{AC}}+S_{\textrm{AD}}+S_{\textrm{BE}}+S_{\textrm{CE}}+S_{\textrm{DE}}+S_{\textrm{BCD}}+S_{\textrm{ABCE}}+S_{\textrm{ABDE}}+S_{\textrm{ACDE}}$ &
	\inneritem $\label{Q3}	3 S_{\textrm{ABC}}+3 S_{\textrm{ABD}}+S_{\textrm{ABE}}+S_{\textrm{ACD}}+3 S_{\textrm{ACE}}+S_{\textrm{ADE}}+S_{\textrm{BCD}}+S_{\textrm{BCE}}+S_{\textrm{BDE}}+S_{\textrm{CDE}} \geq 2 S_{\textrm{AB}}+2 S_{\textrm{AC}}+S_{\textrm{AD}}+S_{\textrm{AE}}+S_{\textrm{BC}}+2 S_{\textrm{BD}}+2 S_{\textrm{CE}}+S_{\textrm{DE}}+2 S_{\textrm{ABCD}}+2 S_{\textrm{ABCE}}+S_{\textrm{ABDE}}+S_{\textrm{ACDE}}$ \\
	\inneritem $\label{Q4} 2 S_{\textrm{ABC}}+S_{\textrm{ABD}}+S_{\textrm{ABE}}+S_{\textrm{ACD}}+S_{\textrm{ADE}}+S_{\textrm{BCE}}+S_{\textrm{BDE}} \geq S_{\textrm{AB}}+S_{\textrm{AC}}+S_{\textrm{AD}}+S_{\textrm{BC}}+S_{\textrm{BE}}+S_{\textrm{DE}}+S_{\textrm{ABCD}}+S_{\textrm{ABCE}}+S_{\textrm{ABDE}}$ &
	\inneritem $\label{Q5} S_{\textrm{AD}}+S_{\textrm{BC}}+ S_{\textrm{ABE}}+S_{\textrm{ACE}}+S_{\textrm{ADE}}+S_{\textrm{BDE}}+S_{\textrm{CDE}} \geq S_{\textrm{A}}+S_{\textrm{B}}+S_{\textrm{C}}+S_{\textrm{D}}+S_{\textrm{AE}}+S_{\textrm{DE}}+S_{\textrm{BCE}}+S_{\textrm{ABDE}}+S_{\textrm{ACDE}} $ \\
\end{rowenumerate}
\vspace{7pt}
\hrule
\end{table}

\begin{table}[b]
	\caption{Representatives for each of the $8$ proto-entropic configurations which generate the information quantities associated to each respective inequality orbit as a primitive of the holographic entropy cone $\mathcal{C}_5$ for $5$ regions. For inequalities \ref{MMI2}$-$\ref{Q5}, the necessary canonical building blocks required to reach rank $D-1$ can be straightforwardly obtained by completing the span of the orthogonal complement of the associated information quantity and are thus omitted for clarity. Here, $\overline{k} \equiv [5]\smallsetminus\{k\}$ for $k\in[5]$ and $\overline{I} \equiv [5]\smallsetminus I$ for $I\subset[5]$. The notation for building blocks is adapted from~\cite{Hubeny2018b}: $\EuScript{C}^\circ[I]$ denotes the canonical building block with a connected surface computing the entropy of $I$, whereas $\EuScript{C}^\ast[I(J)]$ and $\EuScript{C}^\circledast[I(J)]$ refer to the non-canonical building blocks constructed in Sec. $6$ of~\cite{Hubeny2018b} with and without connected surface for $J$, respectively (see Figs. $5(a)$, $5(c)$ and $5(d)$ in~\cite{Hubeny2018b} for respective examples of $\EuScript{C}^\circ$, $\EuScript{C}^\ast$ and $\EuScript{C}^\circledast$).}
	\label{primitives}
	\hrule
	\setlength{\tabcolsep}{4pt}
	\renewcommand{\arraystretch}{2.5}
	\noindent
	\begin{rowenumerate}{J{0.42\textwidth} J{0.53\textwidth} }
		\inneritem{$ \displaystyle \hspace{-6pt} \bigsqcup_{I\in \wp_N \smallsetminus \{\textrm{AB}\}} \hspace{-14pt}\EuScript{C}^\circ[I]$} &
		\inneritem{$ \displaystyle \hspace{-6pt} \bigsqcup_{I\in \wp_N \smallsetminus \{\textrm{ABC}\}} \hspace{-16pt}\EuScript{C}^\circ[I]$} \\
		\inneritem{$ \displaystyle \hspace{-6pt} \bigsqcup_{k\in \{\textrm{B,C,D,E}\}} \hspace{-15pt}\EuScript{C}^\circledast_5[k (\overline{k})] \sqcup \EuScript{C}_4^\ast[\textrm{\small B}(\overline{\textrm{\small BE}})] \sqcup \EuScript{C}_4^\ast[\textrm{\small B}(\overline{\textrm{\small BD}})] $} &
		\inneritem{$ \displaystyle \bigsqcup_{k\in [5]} \EuScript{C}^\circledast_5[k (\overline{k})] \sqcup \EuScript{C}_4^\ast[\textrm{\small A}(\overline{\textrm{\small AE}})] \sqcup \EuScript{C}_4^\ast[\textrm{\small A}(\overline{\textrm{\small AD}})] \sqcup \EuScript{C}_4^\ast[\textrm{\small B}(\overline{\textrm{\small BC}})]$} \\
		\inneritem{$ \displaystyle \hspace{-6pt} \bigsqcup_{k\in \{\textrm{A,B,C,D}\}} \hspace{-15pt}\EuScript{C}^\circledast_5[k (\overline{k})] \sqcup \EuScript{C}_4^\ast[\textrm{\small B}(\overline{\textrm{\small BD}})] $} &
		\inneritem{$ \displaystyle \bigsqcup_{k\in [5]} \EuScript{C}^\circledast_5[k (\overline{k})] \sqcup \EuScript{C}_4^\ast[\textrm{\small B}(\overline{\textrm{\small BE}})] \sqcup \EuScript{C}_4^\ast[\textrm{\small C}(\overline{\small \textrm{CD}})] $} \\
		\inneritem{$ \displaystyle \bigsqcup_{k\in [5]} \EuScript{C}^\circledast_5[k (\overline{k})] \sqcup \EuScript{C}_4^\ast[\textrm{\small A}(\overline{\textrm{\small AE}})] $} &
		\inneritem{$ \displaystyle \EuScript{C}^\circledast_5[ \textrm{\small E} (\overline{\textrm{\small E}})] \sqcup \EuScript{C}_4^\ast[\textrm{\small B}(\overline{\textrm{\small BC}})] \sqcup \EuScript{C}_4^\ast[\textrm{\small C}(\overline{\textrm{\small BC}})] $}
	\end{rowenumerate}
	\vspace{5pt}
	\hrule
\end{table}

It is worth remarking that, as defined, $\mathcal{C}_N$ is the space of holographic entropy vectors for states with time-reflection symmetry to which the RT prescription applies. In principle, lifting this restriction to the fully covariant HRT case could allow for a larger space of entropy vectors, the HRT holographic entropy cone $\mathcal{C}_N^{\textrm{HRT}} \supseteq \mathcal{C}_N$. While the original RT-based proof of strong subadditivity~\cite{Headrick2007} was extended to dynamical setups and proofs of MMI~\cite{Hayden2013,Wall2014}, it has been argued that the same methods may not be generalizable to non-static proofs for the $5$-party inequalities~\cite{Rota2018}. Alternative bit-thread-based proofs of MMI~\cite{Hubeny2018,Cui2018} may lend themselves to generalizations to larger-$N$ inequalities and covariance, but this is yet to be explored. The validity of inequalities \ref{QCyclic}$-$\ref{Q5} for dynamical spacetimes thus remains an open question which has only been verified in specific setups\footnote{In $2+1$ bulk dimensions, \cite{Czech2019} has recently proven that any inequality for which a contraction map exists is not only valid for RT, but also for HRT. In particular, this implies $\mathcal{C}_N^{\textrm{HRT}} = \mathcal{C}_N$ for $N\leq5$ in $2+1$ bulk dimensions.}~\cite{Erdmenger2017,Bao2018b,Caginalp2019}. However, a suggestive indication that $\mathcal{C}_N^{\textrm{HRT}}$ is no larger than $\mathcal{C}_N$ is precisely the algebraic derivation of these from the holographic entropy arrangement, which is defined for arbitrary spacetimes. More importantly, all facets of $\mathcal{C}_5$ can be shown to be primitive quantities as defined in \cite{Hubeny2018a,Hubeny2018b}, thus corresponding to phase transitions of entangling surfaces for arbitrary geometric states. Explicitly, using the proto-entropic formalism and notation for building blocks established in \cite{Hubeny2018b}, Table~\ref{primitives} provides a set of configurations which suffice to generate the information quantities associated to inequalities \ref{SA}$-$\ref{Q5} as primitive, respectively. It is remarkable that, besides canonical building blocks, only the non-adjoining configurations $\EuScript{C}_4^\ast$ and $\EuScript{C}^\circledast_5$ are needed to generate all facets of the polyhedron for $N=5$ up to symmetries (see Table~\ref{primitives} for notation). Note also the necessity of considering non-simply connected boundary topologies with enveloping, for otherwise the $I_n$ theorem would preclude the construction of these quantities as primitive~\cite{Hubeny2018a}. The configurations in Table~\ref{primitives} strongly support the conjecture in~\cite{Hubeny2018b} that the holographic entropy cone and polyhedron are indeed the same object.

\subsection{Extreme Rays}

\begin{table}[t]
	\caption{Representatives for each of the $19$ extreme ray orbits of the holographic entropy cone $\mathcal{C}_5$ for $5$ regions. \hspace*{\fill}}
	\label{extremerays}
	\setlength{\tabcolsep}{6pt}
	\renewcommand{\arraystretch}{1.5}
	\noindent
	\fontsize{9}{10} \selectfont
	\hrule
	\vspace{4pt}
	\begin{rowenumerate}{R{0.47\textwidth} R{0.47\textwidth}}
		\inneritem{$\left(1\x0\x0\x0\x0\x;1\x1\x1\x1\x0\x0\x0\x0\x0\x0\x;1\x1\x1\x1\x1\x1\x0\x0\x0\x0\x;1\x1\x1\x1\x0\x;1\right)$} &
		\inneritem{$\left(1\x1\x1\x0\x0\x;2\x2\x1\x1\x2\x1\x1\x1\x1\x0\x;1\x2\x2\x2\x2\x1\x2\x2\x1\x1\x;1\x1\x2\x2\x2\x;1\right)$}\\
		\inneritem{$\left(1\x1\x1\x1\x0\x;2\x2\x2\x1\x2\x2\x1\x2\x1\x1\x;3\x3\x2\x3\x2\x2\x3\x2\x2\x2\x;2\x3\x3\x3\x3\x;2\right)$} &
		\inneritem{$\left(1\x1\x1\x1\x1\x;2\x2\x2\x2\x2\x2\x2\x2\x2\x2\x;3\x3\x3\x3\x3\x3\x3\x3\x3\x3\x;2\x2\x2\x2\x2\x;1\right)$}\\
		\inneritem{$\left(1\x1\x1\x1\x1\x;2\x2\x2\x2\x2\x2\x2\x2\x2\x2\x;3\x3\x3\x3\x3\x3\x3\x3\x3\x3\x;4\x4\x4\x4\x4\x;3\right)$} &
		\inneritem{$\left(1\x1\x1\x1\x2\x;2\x2\x2\x3\x2\x2\x3\x2\x3\x3\x;3\x3\x4\x3\x4\x4\x3\x4\x4\x4\x;4\x3\x3\x3\x3\x;2\right)$}\\
		\inneritem{$\left(1\x1\x1\x2\x2\x;2\x2\x3\x3\x2\x3\x3\x3\x3\x4\x;3\x4\x4\x4\x4\x5\x4\x4\x5\x5\x;5\x5\x4\x4\x4\x;3\right)$} &
		\inneritem{$\left(1\x1\x1\x1\x1\x;2\x2\x2\x2\x2\x2\x2\x2\x2\x2\x;3\x3\x3\x3\x3\x3\x3\x3\x3\x1\x;2\x2\x2\x2\x2\x;1\right)$}\\
		\inneritem{$\left(1\x1\x1\x1\x2\x;2\x2\x2\x3\x2\x2\x3\x2\x3\x3\x;3\x3\x4\x3\x4\x4\x3\x4\x4\x2\x;4\x3\x3\x3\x3\x;2\right)$} &
		\inneritem{$\left(1\x1\x1\x1\x1\x;2\x2\x2\x2\x2\x2\x2\x2\x2\x2\x;2\x3\x3\x3\x3\x3\x2\x3\x3\x2\x;2\x2\x2\x2\x2\x;1\right)$}\\
		\inneritem{$\left(1\x1\x2\x2\x2\x;2\x3\x3\x3\x3\x3\x3\x4\x4\x4\x;4\x4\x4\x5\x5\x3\x5\x3\x5\x4\x;4\x4\x4\x3\x3\x;2\right)$} &
		\inneritem{$\left(1\x1\x1\x1\x1\x;2\x2\x2\x2\x2\x2\x2\x2\x2\x2\x;3\x3\x2\x3\x3\x2\x3\x2\x3\x2\x;2\x2\x2\x2\x2\x;1\right)$}\\
		\inneritem{$\left(1\x1\x1\x1\x1\x;2\x2\x2\x2\x2\x2\x2\x2\x2\x2\x;3\x2\x3\x3\x3\x3\x3\x2\x3\x2\x;2\x2\x2\x2\x2\x;1\right)$} &
		\inneritem{$\left(2\x2\x2\x2\x3\x;4\x4\x4\x5\x4\x4\x5\x4\x5\x5\x;6\x4\x7\x6\x7\x7\x6\x5\x7\x5\x;6\x5\x5\x5\x5\x;3\right)$}\\
		\inneritem{$\left(3\x3\x3\x3\x3\x;6\x6\x6\x6\x6\x6\x6\x6\x6\x6\x;7\x7\x5\x9\x7\x7\x9\x9\x9\x9\x;6\x6\x6\x6\x6\x;3\right)$} &
		\inneritem{$\left(1\x1\x1\x1\x1\x;2\x2\x2\x2\x2\x2\x2\x2\x2\x2\x;3\x3\x2\x2\x3\x3\x2\x2\x3\x3\x;2\x2\x2\x2\x2\x;1\right)$}\\
		\inneritem{$\left(2\x2\x2\x2\x3\x;4\x4\x4\x5\x4\x4\x5\x4\x5\x5\x;4\x6\x5\x6\x7\x5\x6\x7\x7\x7\x;6\x5\x5\x5\x5\x;3\right)$} &
		\inneritem{$\left(3\x3\x3\x3\x3\x;6\x6\x6\x6\x6\x6\x6\x6\x6\x6\x;5\x9\x7\x9\x9\x7\x7\x9\x9\x7\x;6\x6\x6\x6\x6\x;3\right)$}\\
		\inneritem{$\left(3\x3\x3\x3\x3\x;6\x6\x6\x6\x6\x6\x6\x6\x6\x6\x;7\x9\x5\x7\x9\x7\x9\x9\x9\x7\x;6\x6\x6\x6\x6\x;3\right)$}
	\end{rowenumerate}
	\vspace{5pt}
	\hrule
\end{table}

\begin{figure}[h]
	\begin{subfigure}{.15\textwidth}
		\centering\includegraphics[width=\textwidth]{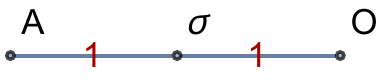}\caption{$\ell_1 = 15.$}
	\end{subfigure}
	\begin{subfigure}{.15\textwidth}
		\centering\includegraphics[width=\textwidth]{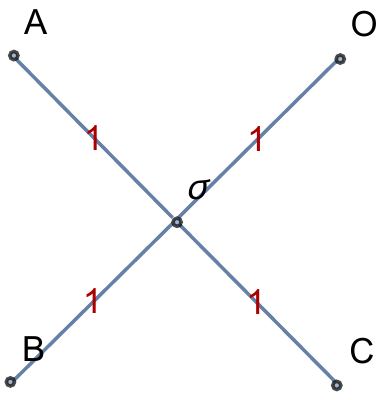}\caption{$\ell_2 = 15.$}
	\end{subfigure}
	\begin{subfigure}{.15\textwidth}
		\centering\includegraphics[width=\textwidth]{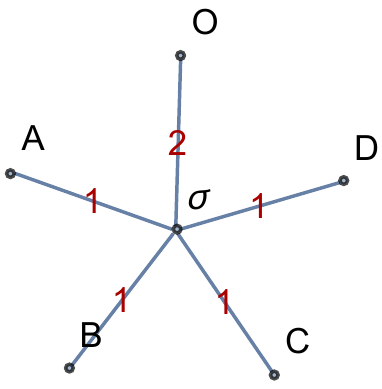}\caption{$\ell_3 = 30.$}
	\end{subfigure}
	\begin{subfigure}{.15\textwidth}
		\centering\includegraphics[width=\textwidth]{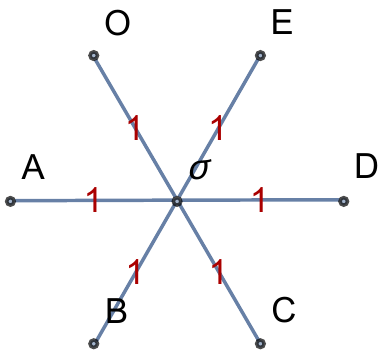}\caption{$\ell_4 = 1.$}
	\end{subfigure}
	\begin{subfigure}{.15\textwidth}
		\centering\includegraphics[width=\textwidth]{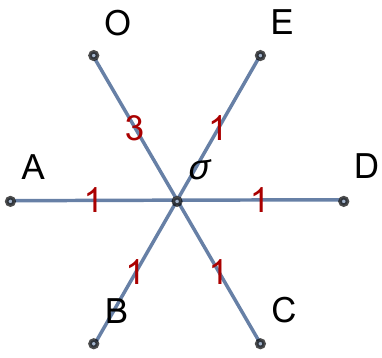}\caption{$\ell_5 = 6.$}
	\end{subfigure}
	\begin{subfigure}{.15\textwidth}
		\centering\includegraphics[width=\textwidth]{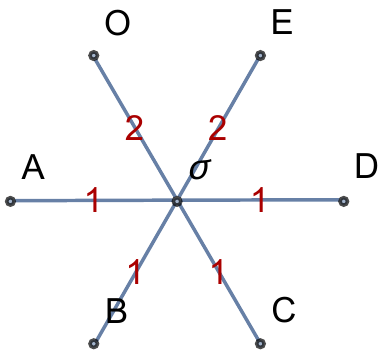}\caption{$\ell_6 = 15.$}
	\end{subfigure}
	\begin{subfigure}{.15\textwidth}
		\centering\includegraphics[width=\textwidth]{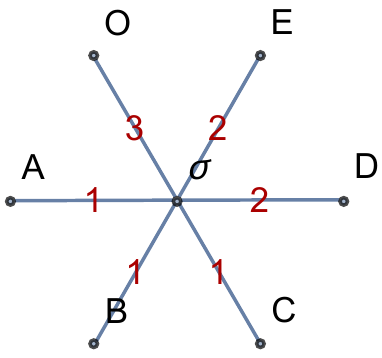}\caption{$\ell_7 = 60.$}
	\end{subfigure}
	\begin{subfigure}{.2\textwidth}
		\centering\includegraphics[width=\textwidth]{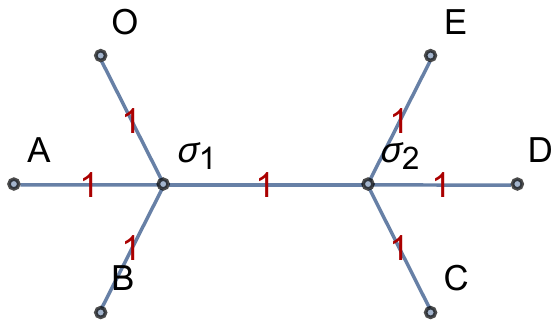}\caption{$\ell_8 = 10.$}
	\end{subfigure}
	\begin{subfigure}{.2\textwidth}
		\centering\includegraphics[width=\textwidth]{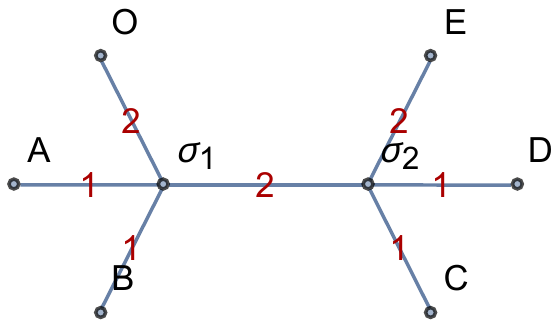}\caption{$\ell_9 = 90.$}
	\end{subfigure}
	\begin{subfigure}{.2\textwidth}
		\centering\includegraphics[width=\textwidth]{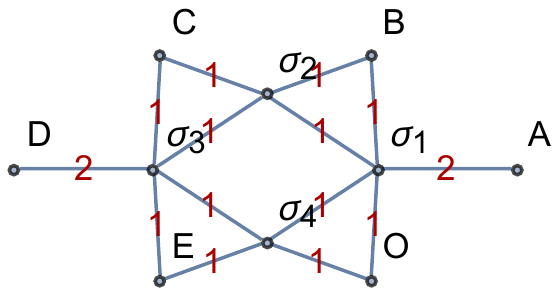}\caption{$\ell_{10} = 60.$}
	\end{subfigure}
	\begin{subfigure}{.2\textwidth}
		\centering\includegraphics[width=\textwidth]{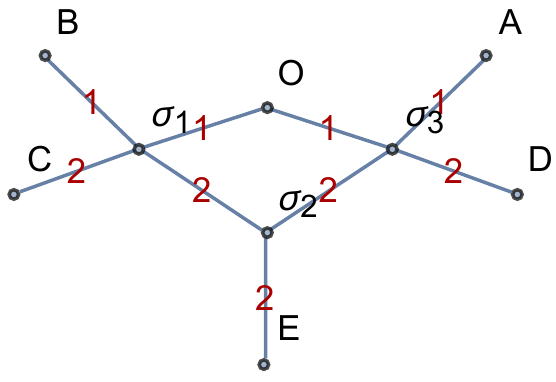}\caption{$\ell_{11} = 180.$}
	\end{subfigure}
	\begin{subfigure}{.2\textwidth}
		\centering\includegraphics[width=.9\textwidth]{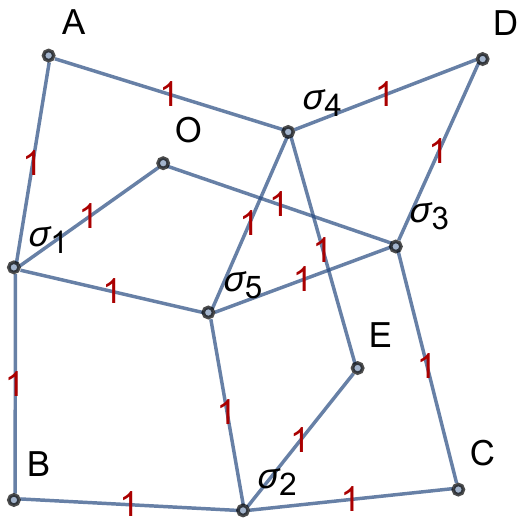}\caption{$\ell_{12} = 15.$}
	\end{subfigure}
	\begin{subfigure}{.25\textwidth}
		\centering\includegraphics[width=\textwidth]{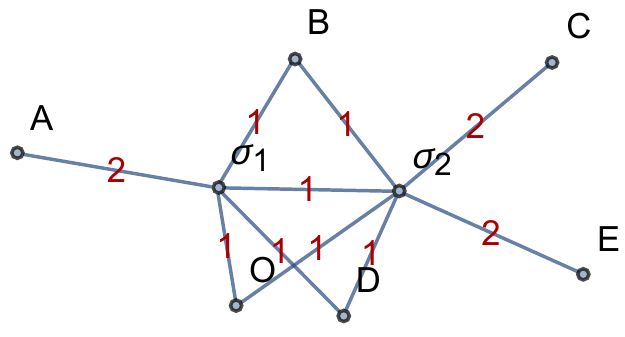}\caption{$\ell_{13} = 60.$}
	\end{subfigure}
	\begin{subfigure}{.25\textwidth}
		\centering\includegraphics[width=\textwidth]{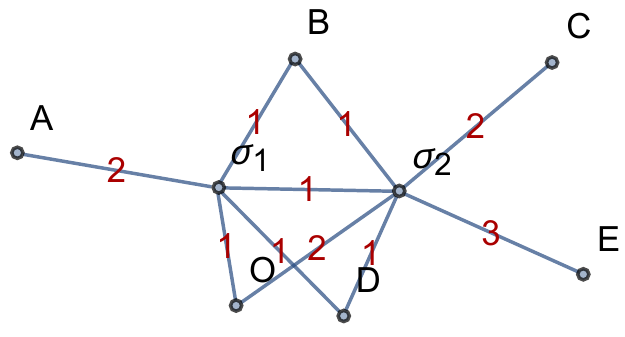}\caption{$\ell_{14} = 360.$}
	\end{subfigure}
	\begin{subfigure}{.25\textwidth}
		\centering\includegraphics[width=\textwidth]{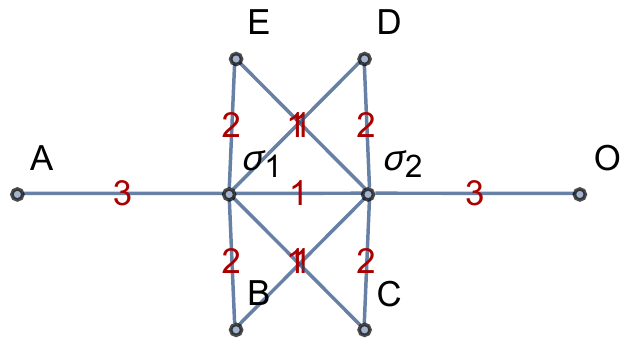}\caption{$\ell_{15} = 90.$}
	\end{subfigure}
	\begin{subfigure}{.24\textwidth}
		\centering\includegraphics[width=\textwidth]{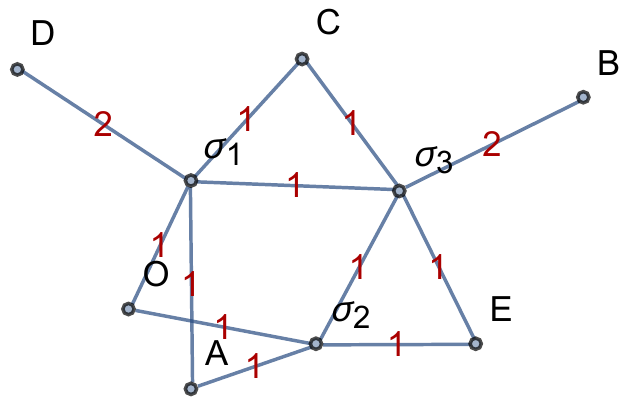}\caption{$\ell_{16} = 180.$}
	\end{subfigure}
	\begin{subfigure}{.2\textwidth}
		\centering\includegraphics[width=\textwidth]{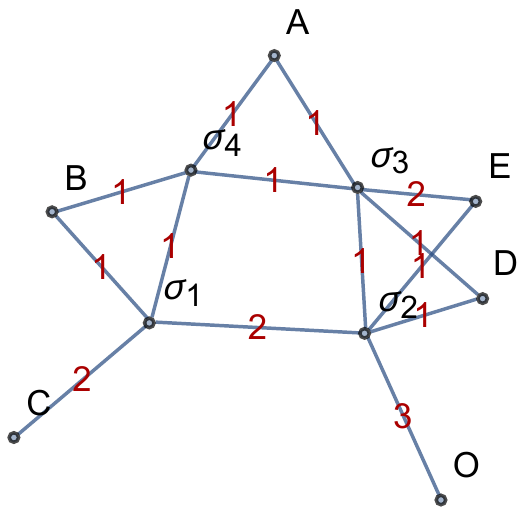}\caption{$\ell_{17} = 360.$}
	\end{subfigure}
	\begin{subfigure}{.25\textwidth}
		\centering\includegraphics[width=\textwidth]{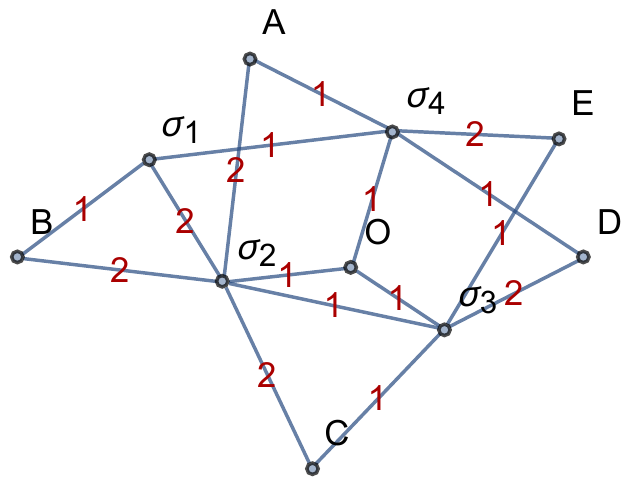}\caption{$\ell_{18} = 360.$}
	\end{subfigure}
	\begin{subfigure}{.25\textwidth}
		\centering\includegraphics[width=\textwidth]{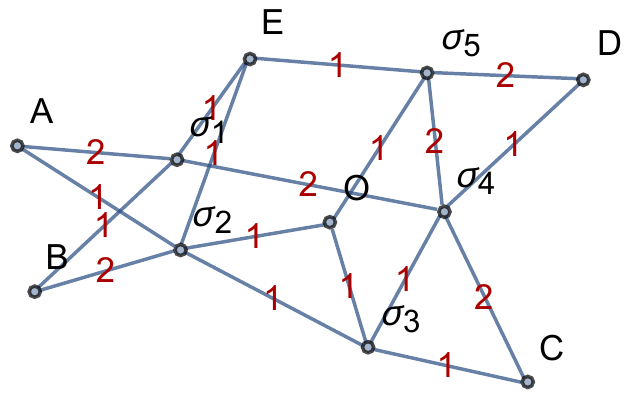}\caption{$\ell_{19} = 360.$}
	\end{subfigure}
	\caption[short]{Graph models realizing the ray representatives in Table~\ref{extremerays}, corresponding to each of the $19$ extreme ray orbits of the holographic entropy cone $\mathcal{C}_5$ for $5$ regions. Graphs are numbered according to the extreme ray they generate, and captioned by the length $\ell_k$ of their orbit. Boundary vertices are labelled by their monochromatic index, bulk vertices by $\sigma_n$, with $n\in\mathbb{Z}^+$ enumerating them, and edges by their capacity. Boundary vertices of pure regions are omitted.}
	\label{extremegraphs}
\end{figure}

The cone specified above by its $372$ facets admits a dual description in terms of $2267$ extreme rays. The latter can be grouped into $19$ distinct symmetry orbits, such that one may focus on a single representative ray per orbit. Table~\ref{extremerays}  shows one such choice of representatives\footnote{Semicolons separate entropy entries corresponding to polychromatic indices of different cardinality. The ordering is given by: ( A, B, C, D, E ; AB, AC, AD, AE, BC, BD, BE, CD, CE, DE ; ABC, ABD, ABE, ACD, ACE, ADE, BCD, BCE, BDE, CDE ; ABCD, ABCE, ABDE, ACDE, BCDE ; ABCDE ).}, while Fig.~\ref{extremegraphs} provides every graph model needed to construct the holographic entropy cone $\mathcal{C}_5$ for $5$ regions\footnote{In producing these graphs, the choice has been to fix the number of boundary vertices to $5+1$, such that the colouring map is bijective for the $5$ regions and their purifier $O$. An explicit construction of an associated wormhole geometry can be accomplished by operations that bring the graph to a suitable form without changing its entropies, as explained in~\cite{Bao2015}.}. The first seven rays continue the pattern of being realizable by star graphs, which prove sufficient to generate all extreme rays for $N\leq4$. However, the other twelve exhibit much richer structure, both in terms of reduced symmetry and non-planarity.

\section{Conclusion}

The holographic entropy cone $\mathcal{C}_N$ is now known for all $N\leq5$. Besides the infinite family of cyclic inequalities, an understanding of the general $N$ case remains elusive. Early explorations of $N=6$ reveal that $\mathcal{C}_6$ consists of at least $19$ valid (i.e. proven by contraction), linearly independent orbits of holographic entropy inequalities. The following is an example of one such $6$-party inequality\footnote{This inequality has been proven using the contraction-map method introduced in~\cite{Bao2015}. It can also be generated as a primitive quantity using the building blocks from~\cite{Hubeny2018b}, thus demonstrating that it is a facet of $\mathcal{C}_6$. The full construction of $\mathcal{C}_6$ is work in progress.}:\\
$S_{\textrm{AB}}+S_{\textrm{ABC}}+S_{\textrm{ACD}}+S_{\textrm{ADE}}+S_{\textrm{BCD}}+S_{\textrm{BDE}}+S_{\textrm{CDE}}+S_{\textrm{CDF}}+S_{\textrm{DEF}}+S_{\textrm{ABCE}} \geq S_{\textrm{A}}+S_{\textrm{B}}+S_{\textrm{AC}}+S_{\textrm{BC}}+S_{\textrm{CD}}+2 S_{\textrm{DE}}+S_{\textrm{DF}}+S_{\textrm{ABE}}+S_{\textrm{ABCD}}+S_{\textrm{CDEF}}+S_{\textrm{ABCDE}}$.

Any constructive approach to exploring $\mathcal{C}_N$ for larger $N$ must overcome the difficulty of dealing with an entropy space of $2^N-1$ dimensions. Already the dual description problem, for which no efficient algorithm is known, can only be feasibly solved up to symmetry~\cite{Bremner}. Moreover, most aspects of the problem suffer a combinatorial explosion which is doubly exponential in $N$ and any hope to proceed constructively must be accompanied by a strategy to tame the combinatorics. In particular, it is indispensable to turn the tables regarding the large degree of redundancy in the structure of $\mathcal{C}_N$ and use its symmetry to one's advantage. Nevertheless, it would ultimately be desirable to understand $\mathcal{C}_N$ for arbitrary $N$. This will most likely require reducing the problem to an algebraic question rather than a combinatorial one, potentially along the lines of the formalism in \cite{Hubeny2018a,Hubeny2018b}.

\addcontentsline{toc}{section}{Acknowledgements}
\acknowledgments

It is a pleasure to thank Xi Dong, Gary Horowitz, Veronika Hubeny, Mukund Rangamani, Massimiliano Rota, Gabriel Trevi\~{n}o Verastegui and Wayne Weng for useful discussions. This research was supported by fellowship LCF/BQ/AA17/11610002 from ``la Caixa'' Foundation (ID 100010434), NSF grant PHY-1801805, and the Center for Scientific Computing from the CNSI, MRL: an NSF MRSEC (DMR-1720256) and NSF CNS-1725797.

%\newpage

\addcontentsline{toc}{section}{References}
%\small{\printbibliography[title={References}]}
\bibliographystyle{JHEPmod} %apsrev4-1, apsrmp4-1, aipauth4-1.bst, aipnum4-1.bst

\bibliography{PaperBib}
%\renewbibmacro*{bbx:savehash}{} %Remove dash from same authors in references

\end{document}